\documentclass{article}

\def\diff{\mathrm d}

\title{Entropy Oriented Trading: A Trading Strategy Based on the Second
Law of Thermodynamics}

\author{Yoichi Hirai\footnote{Department of Information Science, The
Univ. of Tokyo (undergraduate), \texttt{yh@pira.jp}.}}

\begin{document}

\maketitle

\begin{abstract}
The author proposes a finance trading strategy named Entropy Oriented Trading 
and apply thermodynamics on the strategy.  
The state variables are chosen so that the strategy satisfies the 
second law of thermodynamics. 
Using the law, the author proves that the rate of investment (ROI) of 
the strategy is equal to or more than the rate of price change. 
\end{abstract}

\section{Definition of the Strategy}

An Entropy Oriented Trading strategy has $n$ target types of asset
$A_0, \cdots , A_{n-1}$.
It has constant weights for the assets ($w_0, w_1, \cdots , w_{n-1}$), 
where
the weights are positive and the sum of them is one.
The price and the held amount of the $i$th asset are written as $p_i$
and $h_i$ respectively. 
The value of the $i$th asset is 
\[
 U_i = p_i h_i.
\]
The total value of assets is 
\[
 T = \sum_{i=0}^{n-1} U_i. 
\]
The prices $p_i$ change according to time.
On a point of time, a strategy can trade assets $A_i$ and $A_j$.
It can change the held amounts $(h_i, h_j)$ to $(h_i', h_j')$ only if
the total asset is not changed. We define the traded amount of 
value $Q_{ij} = p_i h_i' - p_i h_i$ and express the trade condition 
\begin{equation}
\label{trade}
Q_{ij} + Q_{ji} = 0.
\end{equation}

Multiple trades may occur at the same time. In that case, the sum
$Q_i = \sum_{j=0}^{n-1}Q_{ij}$ is the value of the bought amount 
of the asset $A_i$. The value of $Q_i$ is negative if 
asset $A_i$ is sold.

An Entropy Oriented Trading strategy tries to keep the distribution of
asset equal to the weights, 
so in  equilibrium, the equation (\ref{state}) hold: 
\begin{equation}
\label{state}
 U_i = w_i T.
\end{equation}

When the price change is slow enough (or the strategy trades 
frequently enough), the equation (\ref{state}) holds. 
Otherwise, $T_i = U_i / {w_i}$ can be different from $T$.
In this case, the strategy trades to make all $T_i$'s equal.

We can assume that when $T_i > T_j$, an Entropy Oriented Trading
 strategy does not trade 
selling $A_j$ and buying $A_i$.  This means 
\begin{equation}
 \label{second}
{\rm if}\quad T_i > T_j, \quad Q_{ij} \le 0.
\end{equation}
 A trade involving $A_i$ and $A_j$ is
reversible only if $T_i = T_j$. 

\section{Analysis of the Strategy}

To analyse this strategy with thermodynamics, we have to choose the state
variables and the equation of state. The state variables are $p_i, h_i,
T_i$ and the equation of state is 
\begin{equation}
\label{eos}
 w_i T_i = p_i h_i
\end{equation}
and we take the internal energy as $U_i = w_i T_i$. Note that the
$T$-based 
trading restriction (\ref{second}) is analogous to the second law
of thermodynamics.

We first analyse the case in which the equation (\ref{state}) always holds
and then analyse the more general case.
In both cases, we assume that we can consider a transformation of system
as a continuous sequence of infinitesimal transformations.

\subsection{On Reversible Transformations}

When the equation (\ref{state}) always holds, $T_i
= T_j$ holds for any $i$ and $j$, so that all trades are reversible.
The equation of state becomes
\begin{equation}
\label{eos_reversible}
 w_i T = p_i h_i.
\end{equation}

Taking the differential of (\ref{eos_reversible}), we have 
\[
 w_i \diff T =  h_i\,\diff p_i + p_i\,\diff h_i.
\]
The amount $p_i \, \diff h_i = \diff'Q_i$ is the change of $U_i$ 
due to the trades occurring at the given point of time, whereas the
amount $h_i\,\diff p_i = w_i T {\diff p_i}/{p_i}$ is the change of
the value of $U_i$ due to the 
price change.  

Integrating this from the state $C$ to the state $D$ yields:
\[
 w_i \int_C^D \frac {\diff T} T = w_i \int_C^D
  \frac {\diff p} p + \int_C^D \frac{\diff' Q_i} T.
\]
Now we take the summation for $i$.
For simplicity, we introduce weighted geometric average of prices $P =
\prod_{i=0}^{n-1} p_i^{w_i}$.
Since the trade condition (\ref{trade}) states $\sum_{i=0}^{n-1}
\diff' Q_i / T = 0$, we can omit the term $\int_C^D {\diff' Q_i} / T$ 
and get
\[
 \ln \left(\frac{T(D)}{T(C)}\right) = \ln \left(\frac{P(D)}{P(C)}\right),
\]
stating that the rate of investment is equal to the price change rate.

We take an arbitary equilibrium state $O$ and define the entropy $S_i(C)$
as
\begin{equation}
\label{entropy}
 S_i(A) = \int_O^A \frac{\diff'Q_i}{T_i},
\end{equation}
where the integral is taken along a reversible transformation. 
We also define the total entropy
$S(A)=\sum_{i=0}^{n-1} S_i(A).$
Note that the total entropy $S$ is constant in reversible transformations.

\subsection{On Irreversible Transformations}

In irreversible transformations, the amount $T_i$ can be different for
different $i$ because of delay of trades.  
Taking the differential of the equation of state
(\ref{eos}) yields
$ w_i \diff T_i = p_i\,\diff h_i + h_i\,\diff p_i$. 
Dividing this by $T$ and taking the integral from a state $C$ to 
another state $D$, we get
\[
 w_i \ln\left(\frac{T_i(D)}{T_i(C)}\right) = w_i \ln
  \left(\frac{p_i(D)}{p_i(C)}\right) + \int_C^D\frac{\diff'
  Q_i}{T_i}.
\]

We assume that the states $C$ and $D$ are in  equilibrium so that for any
$i$ $T_i(C) = T(C)$ and $T_i(D) = T(D)$.  
We take summation for $i$ and see
\begin{equation}
\label{fin}
 \ln\left(\frac{T(D)}{T(C)}\right) = \ln\left(\frac{P(D)}{P(C)}\right) +
  (S(D) - S(C)).
\end{equation}

The trading restriction
(\ref{second}) means $Q_{ij}/T_i + Q_{ji}/T_j \ge 0$. From this we can
obtain $\Delta S = S(D) - S(C) \ge 0$. 
Especially, when a irreversible trade occurs between $A_i$ and $A_j$, 
the entropy increases by 
$Q_{ij} / T_i + Q_{ji} / T_j > 0.$

The equation (\ref{fin}) states that the return of investment 
is the product of the rate of price change and $\exp(\Delta S)$, which
is equal to or more than one:
\[
 \Delta(\ln T) = \Delta(\ln P) + \Delta S, \qquad {\rm where}\quad
 \Delta S\ge 0.
\]

\end{document}